\documentclass[reprint,superscriptaddress,prb,longbibliography]{revtex4-2}
\usepackage[utf8]{inputenc}
\usepackage[T1]{fontenc}
\usepackage{amsmath}
\usepackage{amssymb}
\usepackage{graphicx}
\usepackage[capitalize, nameinlink]{cleveref}
\usepackage{subfigure}
\graphicspath{{figures/}}
\usepackage{adjustbox}
\usepackage{booktabs}
\usepackage{multirow}

\usepackage{color}
\usepackage{rotating}
\usepackage{ulem}
\usepackage{color}

\begin{document}

\title{Frequency comb in a macroscopic mechano-magnetic artificial spin ice}

\author{Renju R. Peroor}
\author{Lawrence Scafuri}
\author{Dmytro A. Bozhko}
\author{Ezio Iacocca}
\affiliation{Center for Magnetism and Magnetic Nanostructures, University of Colorado Colorado Springs, Colorado Springs, CO, USA}

\date{\today}

\begin{abstract}
Artificial spin ices are metamaterials composed of interacting nanomagnets exhibiting frustration. Their resonant magnetization dynamics have been broadly investigated from fundamental and applied points of view. In this work, we realize a dynamically driven macroscopic mechano-magnetic artificial spin ice, or macro-ASI, where permanent magnets are allowed to rotate on specially designed hinges and exhibit natural resonance frequencies on the order of several Hertz. A nonlinear dynamical regime is achieved experimentally and well reproduced by numerical modelling. The modulation of the magnetic coupling leads to a frequency comb that manifests itself as an amplitude-phase modulation of the magnets' motion due to a metastable condition, i.e., a Hopf bifurcation. Our results not only demonstrate a striking similarity across different physical systems, but also suggest that the mechanism to enable nonlinear phenomena could be realized in nanoscale systems using microresonators decorated with magnetic materials to dynamically modulate their coupling.
\end{abstract}

\maketitle

Nonlinearities in dynamical systems give rise to a variety of interesting physical effects, many of which have the potential to be harnessed for technological applications. One of the most well-known examples is the generation of an optical frequency comb~\cite{Fortier2019}, in which a coherent and short optical laser pulse train results in a discrete spectrum comprising millions of optical modes. Applications of frequency combs include precision metrology and the development of versatile spectrometers. In the context of communications, compact and phase-synchronous frequency combs can be produced in microresonators taking advantage of Kerr nonlinearity~\cite{Pfeifle2014} or Pockels effect~\cite{Rueda2019} which self-modulate an electromagnetic wave. More recently, frequency combs emerged in micro-mechanical resonators~\cite{Ochs2022,deJong2023} due to their strong nonlinearity~\cite{Defoort2021,Samanta2022}. Nonlinearity can be also present when distinct particles and quasiparticles~\cite{QuasiparticleFeature2016} interact. This has led to the realization of frequency combs utilizing, e.g., acoustic and superconducting resonators~\cite{Han2022} and magnon-phonon interactions with a Yttrium Iron Garnet (YIG) sphere in a cavity~\cite{Xu2023}. Magnetic materials are intrinsically nonlinear so that the quanta of angular momentum, the magnons, have been shown to exhibit frequency combs when driven to nonlinear interactions~\cite{Hula2022} or when interacting with solitons~\cite{Wang2021}.

Another class of materials where frequency combs could be realized are artificial spin ices~\cite{Skjaervo2020}. These are geometrically arranged nanomagnets that host a wealth of dynamical modes~\cite{Jungfleisch2016,Gliga2020,Lendinez2021} 
including nonlinear dynamics~\cite{Lendinez2023}. Many other configurations are possible~\cite{Skjaervo2020}, including extension to three-dimensional lattices~\cite{Alatteili2024}, which have also shown interesting dynamics such as 
coherent waves in tripod-based ASIs~\cite{Sahoo2021,Harding2024} and ultra-strong magnon coupling in 3D square ASIs~\cite{Dion2024}. However, in all these cases, nanoscopic arrangements are spatially fixed due to the lithography processes used, and the GHz dynamics are coupled through edge modes in which the magnetic volume and the concomitant dynamic stray field is greatly reduced in extent. Therefore, frequency combs are likely unavailable in ASIs unless additional degrees of freedom are allowed.

Here, we demonstrate that a frequency comb can emerge from a driven macroscopic mechano-magnetic artificial spin ice system~\cite{Mellado2012,Teixeira2024}, or macro-ASI. By using a macro-ASI, we ensure a strong coupling mediated by stray fields throughout the array. In addition, the mechanical motion of the magnets results in a temporal or self-Floquet modulation of the dipole-dipole interaction. When driven to nonlinear resonance, the system enters an unstable region characterized by a nonlinear amplitude and phase modulation that establishes the comb spectrum. A notable feature of our macro-ASI system is the richness of nonlinear dynamics at macroscopic scales and frequencies perceivable by the naked eye. Our results suggest that frequency combs can be also achieved in geometrically positioned arrays of micro-mechanical resonators with high quality factors (e.g.~\cite{Sun2011,Ochs2022}) decorated with magnetic materials.
\begin{figure*}[t]
\begin{center}
\includegraphics[width=6in]{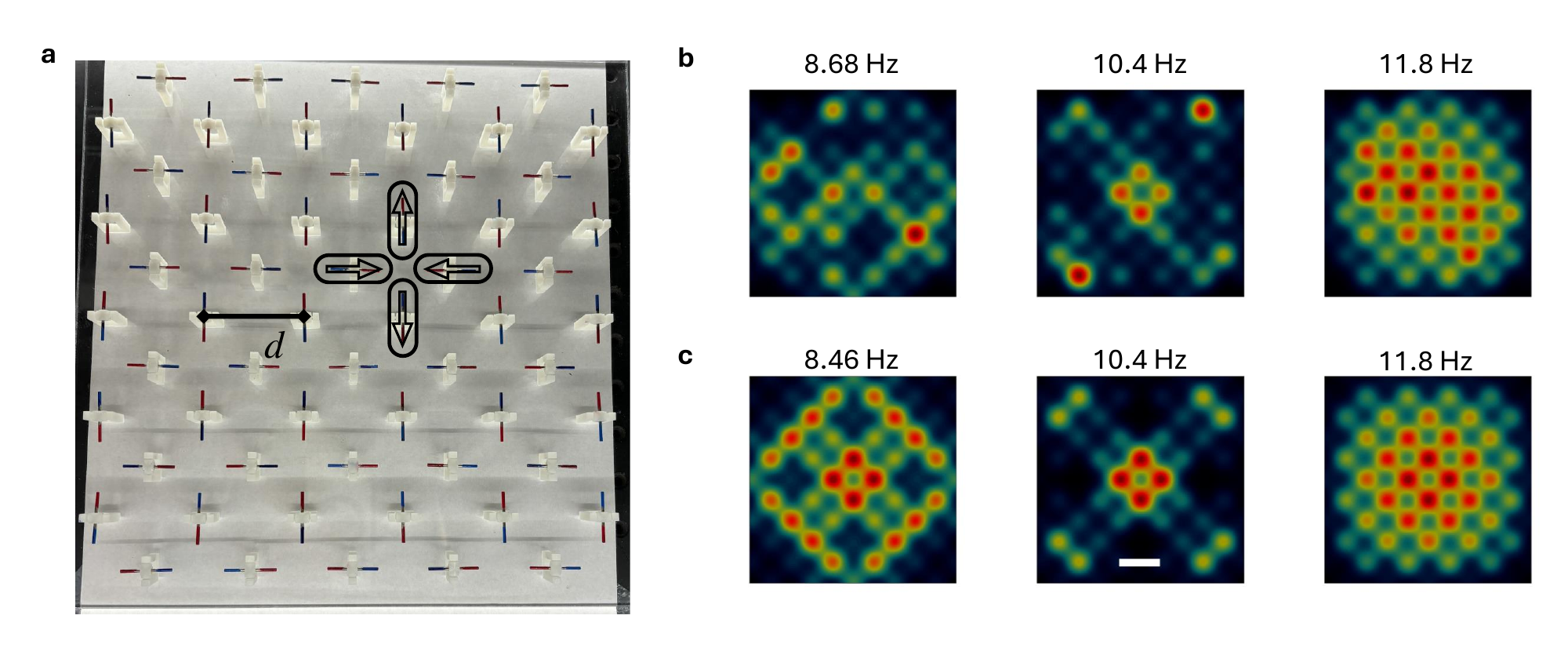}
\caption{\textbf{Resonant modes in a square macro-ASI.} \textbf{a} Picture of the macro-ASI composed of 60 magnets arranged in a square lattice. Resonant modes obtained from \textbf{b} experiments and \textbf{c} numerical modeling. The modes are in good quantitative agreement, both in their spatial symmetries and resonant frequencies. The scale bar represents $5$~cm.}
\label{fig1}
\end{center}
\end{figure*}

A geometric arrangement is crucial to obtain a strongly coupled system. We used the square ASI configuration~\cite{Wang2006} shown in Fig.~\ref{fig1}\textbf{a}. The finite macro-ASI lattice is composed of 2.54-cm-long cylindrical permanent magnets supported by rotary hinges with low friction. The lattice constant of this square configuration is $d=5.08$~cm. Four magnets strongly interact at each vertex composing a unit cell that is schematically illustrated in Fig.~\ref{fig1}\textbf{a} with arrows towards each magnet's north pole. The system immediately achieves its ground state or type-I configuration which conforms with the ``ice-rule''~\cite{Gilbert2014}. Because of the geometry, a square ASI configuration exhibits only short-range frustration, i.e., any defect can be compensated and large sections of the array are defect-free~\cite{Wang2006}. In fact, our finite macro-ASI composed of 60 permanent magnets displayed a uniform ground state. Details on the macro-ASI fabrication are given in the Methods section. In contrast to previous realization of macroscopic ASIs~\cite{Novak2004,Mellado2012,Teixeira2024}, our set-up was built with a focus on driven dynamics rather than population of defects due to frustration from a high-energy state. Additional attention was paid to achieve an accurate lattice geometry and uniformity of the mechanical properties of individual elements. As a result, the macro-ASI exhibits wave propagation due to a localized perturbation, as shown in the Supplementary Video 1~\cite{OSF}.

Because of its finite size, the system supports only discrete standing wave modes of oscillation. These are experimentally measured by perturbing the ground state with a magnetic field that approximates a Heaviside function and capturing the time-dependent dynamics of each magnet's rotation angle $\phi$ after the field is rapidly turned off. The magnetic field was produced by a solenoid of radius $a=10.5$~cm placed directly under the macro-ASI. The experimental details on the excitation as well as the data acquisition and processing is described in the Methods. Fourier analysis is performed on each time trace to obtain the frequency-dependent amplitude as a function of space. The reconstructed standing wave patterns are shown in Fig.~\ref{fig1}\textbf{b}, where the Fourier peak amplitude for each magnet was used to scale a two-dimensional Gaussian function for better visualization. The symmetry of the system is immediately apparent from the resulting standing wave patterns. The modes at $8.68$~Hz and $10.4$~Hz both exhibit excitation at the center of the array and symmetric excitation of magnets towards the diagonal edges of the array. The mode at $11.8$~Hz corresponds to a ``bulk'' mode.

\begin{figure*}[t]
\begin{center}
\includegraphics[width=6in]{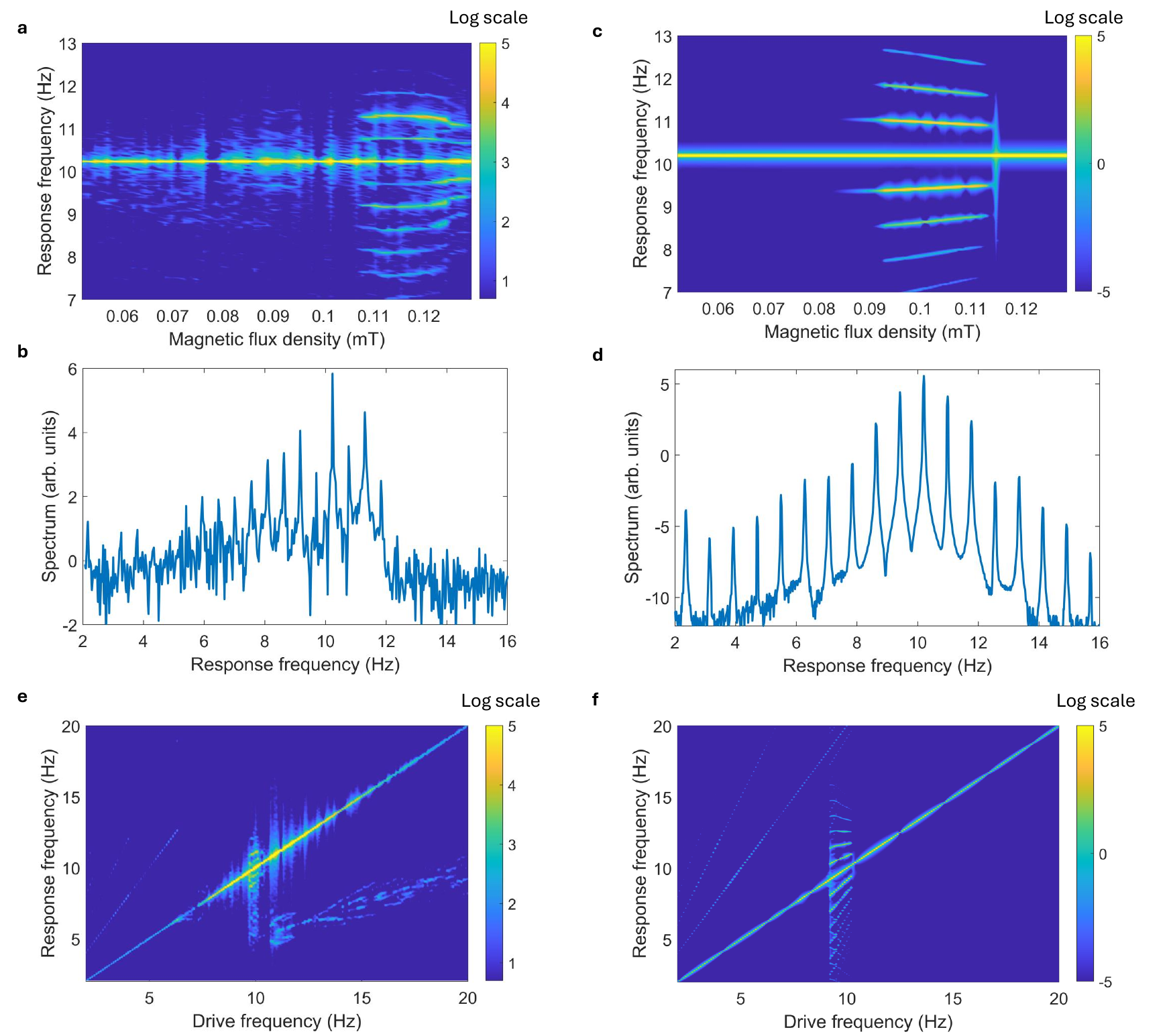}
\caption{\textbf{Observation of frequency comb.} The experimental measurement of frequency combs is shown in \textbf{a}, \textbf{b}, and \textbf{e}. \textbf{a} Magnetic spectrum showing the emergence of a comb as a function of the magnetic flux density of the drive at $10.2$~Hz. \textbf{b} Line-cut of the comb spectrum in \textbf{a} at $B=0.112$~mT. \textbf{e} Drive frequency dependence at $B=0.13$~mT showing the comb in the vicinity of 10~Hz. The numerical modeling of frequency combs is shown in \textbf{c}, \textbf{d}, and \textbf{f}. \textbf{c} Magnetic spectrum showing the emergence of a comb as a function of the magnetic flux density of the drive at $10.2$~Hz. \textbf{d} Line-cut of the comb spectrum in \textbf{c} at $B=0.1$~mT. \textbf{f} Drive frequency dependence at $B=0.1$~mT showing the comb in the vicinity of 10~Hz. }
\label{fig2}
\end{center}
\end{figure*}

These standing wave patterns are well reproduced by a model that considers the torque on each magnet due to the dipole-dipole interaction between all magnets as well as mechanical friction and inertia~\cite{Mellado2012,Teixeira2024}. Details on the model and its implementation are given in the Methods. The resulting standing wave patterns are shown in Fig.~\ref{fig1}\textbf{c}. We obtained a remarkable agreement between experiment and modeling for all standing wave modes, see the Supplementary Figure~\ref{SI:fig1}.

The harmonically-driven dynamics of the macro-ASI were achieved by passing 
an alternating current produced by a function generator through the solenoid, as described in the Methods. The field-magnitude-dependent spectrum at a drive frequency of $f_\mathrm{exc} = 10.2$~Hz is shown in Fig.~\ref{fig2}\textbf{a}, exemplary for one magnet of the array. These results are consistent for all magnets, as shown in the Supplementary Figure~\ref{SI:fig2}. At low flux density magnitudes, only the direct excitation is observed, indicating a linear dynamical regime. However, a change is observed at the threshold value of $B_1=0.104$~mT, above which the spectrum displays equidistant harmonics. This feature indicates that the dynamics become nonlinear, resulting in the emergence of a frequency comb. For example, the spectrum at $B=0.112$~mT is shown in Fig.~\ref{fig2}\textbf{b} for an extended frequency range. Up to thirteen peaks can be clearly discerned between 5~Hz and 12~Hz.  Note, that the optical measurement method utilized in this study is linear with respect to the magnet's angle and does not distort the measured dynamics as an electromagnetic pickup coil would.

The onset of the frequency comb depends also on the drive frequency. The frequency-dependent spectrum at our maximum flux density of $B=0.13$~mT is shown in Fig.~\ref{fig2}\textbf{e}. The comb is only visible in the vicinity of $f_\mathrm{exc}\approx10$~Hz. Over this drive frequency, harmonic content about $f_\mathrm{exc}/2$ appears with well-defined features. The appearance of these half-frequency excitations in the spectrum could be understood as a result of a parametric instability of the forced oscillations into a pair of standing wave modes, which exist in our system at the half of the drive frequency (see Supplementary Figure~\ref{SI:fig1}). Further investigations of this interesting phenomenon requires more detailed insight into the dispersion relation of propagating waves in this system and is outside of the scope of this particular work. We note that these spectra have a striking resemblance to those obtained numerically for a nanoscopic square ASI driven to nonlinear regime~\cite{Lendinez2023}, underscoring the relevance of mode mixing. However, in nanoscopic ASIs, the arrays are extensive and modes are located within the nanomagnets whereas in the present macro-ASI the modes are standing waves resulting from the constrained dimensions of the array.

\begin{figure*}[t]
\begin{center}
\includegraphics[width=6in]{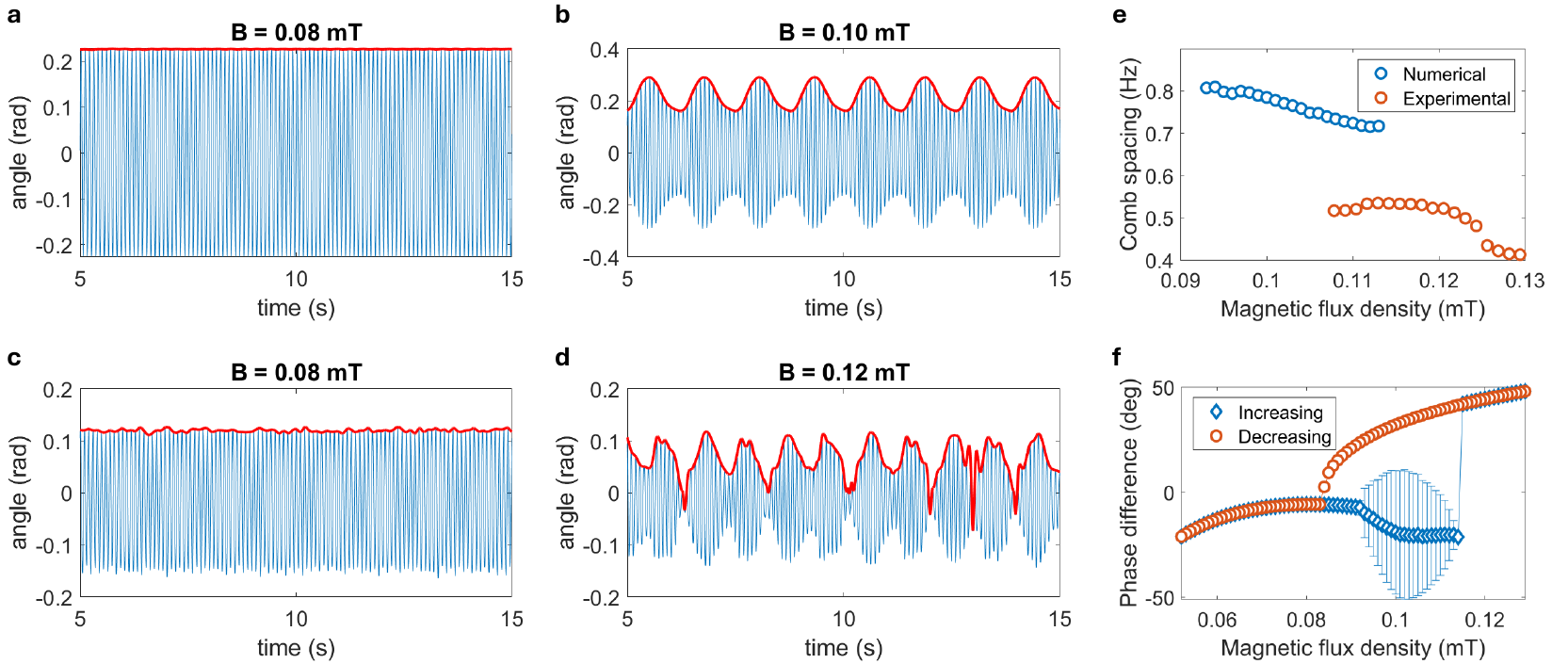}
\caption{\textbf{Characteristics of frequency combs.} Numerical modeling traces of the angle excursion in the \textbf{a} linear and \textbf{b} nonlinear regimes. Experimental traces of the angle excursion in the \textbf{c} linear and \textbf{d} nonlinear regimes. The nonlinear amplitude modulation in \textbf{b} and \textbf{d} lead to the comb spectrum. \textbf{e} Comb spacing obtained from numerical and experimental spectra. In both cases, the comb spacing decreases as a function of magnetic flux density. Errorbars represent the standard deviation in the comb frequency determination by a fitting procedure. \textbf{f} Phase difference between the nanomagnet trace and the drive frequency when the magnetic flux density increases (blue circles) and decreases (red diamonds). The errorbars represent the maximum phase difference excursion. The comb occurs when the phase is modulated due to a Hopf bifurcation.}
\label{fig3}
\end{center}
\end{figure*}

To gain further understanding of the behavior of the system, we performed numerical modelling, which reproduced well the experimentally observed driven dynamics. To capture possible effects emerging from non-uniformity of the excitation field, we considered the spatially-inhomogeneous field distribution due to a solenoid aligned with the geometric centre of the macro-ASI, as described in the Methods. For the given experimental conditions, we obtain the spectrum dependent on the magnetic flux density shown in Fig.~\ref{fig2}\textbf{d}. Similar to the experiments, the model exhibits a comb spectrum for a range of flux density magnitudes. The transition from linear to comb is rather smooth and occurs at similar flux density magnitudes, on the order of 0.1~mT. The lack of imperfections in the numerical model leads to a large number of harmonics in the comb. This is further evidenced by the frequency-dependent spectra for $B=0.01$~mT shown in Fig.~\ref{fig2}\textbf{e}. As a function of drive frequency, the comb is also observed in a narrow band of frequencies, as shown in Fig.~\ref{fig2}\textbf{f}. However, the modeling does not exhibit the half-frequency harmonic content at diving frequencies over $10$~Hz. As the physical origins of the secondary half-frequency dynamics are understood in the frame of parametric processes, we will leave further refinements of the numerical model to include imperfections and vibrations for a future work.

To elucidate the origin of the comb spectrum, we analyse the time-traces of the tilt angle, $\phi$. For consistency, we investigate the same magnet used for Fig.~\ref{fig2}. Below the threshold, both modelling and experiment return an ideal sinusoidal variation, shown in Fig.~\ref{fig3}\textbf{a} and \textbf{b}, respectively. The envelope is emphasized by a solid red curve. While the experiment clearly demonstrates the impact of noise, the envelope is essentially constant. At fields above threshold, the situation dramatically changes. The envelope is a nonlinear function which modulates the driven dynamics, shown in Fig.~\ref{fig3}\textbf{c} and \textbf{d} from modelling and experiment, respectively. The modulation period is observed to be on the order of 1 second. The fact that the modulation is nonlinear is important as this leads to a comb spectrum rather than a simple amplitude modulation from a pure tone. The Supplementary Video 2~\cite{OSF} demonstrates that the amplitude modulation is clearly visible in the experiment.

The frequency spacing corresponds to the fundamental frequency of the amplitude modulation. The extracted comb spacings are shown in Fig.~\ref{fig3}\textbf{e}. While there is a clear quantitative disagreement between the numerical calculations and the experiments, the comb spacing is estimated to be in the close range of 0.4 to 0.8~Hz, as expected from the modulation period of the time traces. To accurately determine the comb frequency spacing, we fitted the resonant frequency and six comb sidebands with Lorentzian functions, resulting in errors under $6$~mHz and $12$~mHz from modelling and experiment, respectively. It is also apparent that the comb spacing decreases as a function of the drive magnetic flux density. This effect can be understood as a consequence of an increased amplitude of oscillation. Such an increase correlates with effective distance between poles of magnets, which in turn leads to an effective reduction in the coupling strength between them that makes modulation frequency smaller.

To further clarify the mechanism giving rise to the comb spectrum, we extracted the relative phase between the magnet and the drive frequency from the modelling. This is achieved by computation of the phase difference between the drive signal $b$ and the angle $\phi$~\cite{Pikovsky2001}
\begin{equation}
    \Delta \theta = \tan^{-1}\left(\frac{b^*\phi-b\phi^*}{b\phi+b^*\phi^*}\right),
\end{equation}
at each driving magnetic flux density and where the superscript $*$ denotes here the Hilbert transform. The resulting phase difference as a function of increasing flux density magnitude is shown in Fig.~\ref{fig3}\textbf{f}, where the blue diamonds represent the average phase difference and the errorbars represent the maximum and minimum phase difference. In this representation, visible errorbars indicate the appearance of the frequency comb. This implies that the comb spectrum results from a nonlinear phase and amplitude modulation of the temporal dynamics.

When the field decreases in magnitude, shown by red circles, the comb is absent. We now understand the comb as a result of a Hopf bifurcation in the system: the fixed point solution transitions into a periodic orbit. Physically, the bifurcation occurs because the drive frequency excites coexisting modes with different phases. As a result, the magnets periodically transition from one mode to the other. This is similar to the process observed in a nanomechanical resonator~\cite{Ochs2022} in terms of nonlinear resonance and also as frequency pulling~\cite{Shoshani2023}. However, our system is driven into nonlinearity due to the strong dynamic coupling between the magnets. In fact, the model shows that at fields over $B\approx1.02$~mT, linearity is restored due to the dynamic mode coupling becoming negligible. The mode metastabiliy can be observed experimentally by initiating the dynamics without any particular protocol, as evident from the Supplementary Videos 2 and 3~\cite{OSF}.

Our findings are in stark qualitative contrast to the magnetisation dynamics of nanomagnets, which are affixed to a surface and are only weakly coupled through edge modes, even in the nonlinear regime~\cite{Lendinez2023}. This suggests the possibility to achieve richer dynamics in nanomagnets patterned on micro-mechanical oscillators, enabling dynamic coupling through the magnet's strong static stray field and its modulation through high quality factor mechanical motion. Such devices would achieve a nonlinear regime at smaller driving fields due to the strong magnetic coupling between elements, possibly even in a quantum regime. From a broader perspective, our results demonstrate the generality of nonlinear phenomena giving rise to frequency combs, including macroscopic scales.

\section*{Methods}
\subsection*{Macro-ASI fabrication}

The macro-ASI was based on N42 type 1-inch (25.4~mm) long and 1/16-inch (1.59mm) diameter cylindrical permanent magnets purchased from K\&J Magnetics~\cite{KJMagnetics}. To quantify magnetic properties of these magnets we performed additional magnetisation measurements using a SQUID magnetometer, described in the Supplementary Note~\ref{SI:SQUID}. The north and south poles were painted for proper identification of the ground state of the array. Each magnet was attached to a rotary hinge that was 3D printed with an ANYCUBIC Photon Mono X Stereolithography (SLA) resin printer with $50\,\mathrm{\mu m}$ spatial resolution. The magnets were introduced into the rotors prior to UV-light curing, so that the magnets were tightly affixed to the rotor. The rotor was then placed on a 3D-printed U-shaped support, composing the macro-ASI lattice element. The produced magnet assemblies allow magnets to freely rotate about the pitch axis and experience negligible yaw as well as low friction due to fine spatial resolution of the 3D printer. Each U-shape support was designed to have two 1~mm long and 1~mm diameter alignment pins at their bottom for accurate positioning on a surface of a assembly board.

The 60-magnet square macro-ASI was built on a $12$~in~$\times12$~in acrylic base. A precise location of the macro-ASI elements was achieved by matching the alignment pins of the U-shaped supports with 1~mm holes drilled using precision Genmitsu 3018-PRO CNC desktop mill (improved in-house to achieve $\approx10\,\mathrm{\mu m}$ accuracy).

\subsection*{Experimental data acquisition and processing}

The experiment was set up by driving forced oscillations using an external magnetic field. To achieve this, we introduced a magnetic coil with an outer diameter of $22$~cm and an inner diameter of $19$~cm. The coil was positioned beneath the acrylic base where the magnet holders are aligned and glued. Both coil and acrylic base were mounted on a rotating stand. The coil was driven by a Siglent SDG2122X arbitrary waveform generator. We used a high-speed camera with frame rate of 250 frames per second to capture the magnets' dynamics with sufficient oversampling to achieve better signal to noise ratio. We used a FLIR camera connected to a tele-objective lens with a focal length ranging from 12.5~mm to 75~mm placed at a distance of approximately 1.5~m from the array to minimize errors of magnet tilt determination  resulting from perspective distortions. The experiment was performed in two stages. First, the array was aligned to capture the 30 magnets parallel to one side of the acrylic base. Then, the stand was rotated 90 degrees to capture the dynamics of the 30 remaining magnets. Because of only one camera available for the acquisition, we limited ourselves to time-domain measurements followed by the spectral domain data evaluation, while the phase dynamics was evaluated using numerical simulations. 

The solenoid has a resistance of approximately 1~Ohm and to be safely directly driven by the arbitrary waveform generator, a 50~Ohm impedance-matching resistor was introduced in series with the coil. Therefore, the maximal achievable amplitude of driving voltage was limited to 10.196~V. To determine the proportionality between the voltage and the field produced by the solenoid, we used a DC current source to directly drive the solenoid at a range of currents from $0$ to $3$~A and measured the resulting field at a position aligned with the solenoid's centre and at the height of the magnetic array. At that position and at $1$~A, the field was $6.6\times10^{-4}\pm0.2\times10^{-4}$~T. Considering the maximal voltage and the total resistance of the circuit of $51\,\mathrm{Ohm}$ the scaling factor for field is thus $1.29\times10^{-4}~$T/V.

The frequency of the driving harmonic voltage was systematically swept from 2.0~Hz to 20.0~Hz. To ensure the system was stabilized, we used to following protocol: first, the frequency was linearly swept from 0~Hz to the desired measurement frequency in $30$~s. Then the system was let to stabilize for $10$~s. Finally, the measurements were taken by capturing a $30$~s AVI video with the camera.

After the completion of the recording of the entire frequency sweep, the data was processed using a specially developed Python script based on OpenCV library \cite{OpenCV}. Firstly, bounding rectangles containing each of the 30 magnets were cropped individually and then the blue and the red binary color masks were applied. Then, a line weighted fit connecting the pixels was applied to estimate the magnet orientation. The deviation of the fitted line from the horizontal axis indicated the angle of the magnet per frame. Each of these traces was analysed by fast Fourier transform (FFT).

\subsection*{Numerical modeling}

Each magnet was modelled as two magnetic ``monopoles'' carrying a magnetic charge: \(+q\) for the north pole and \(-q\) for the south pole~\cite{Mellado2012,Teixeira2024}. The monopole ``charge'' is defined as $q\equiv M_s\pi D^2/4$, where $D$ is the diameter of the magnet and has units of A~m. We consider interactions among magnets governed by a Coulomb-like force between monopoles throughout the array, interaction with an external magnetic flux density, and energy dissipation due to friction at the hinges of the rotors. These give rise to an equation of motion whereby the torque experienced by monopole $i$ is:
\begin{widetext}
    \begin{equation}
    \label{eq:SItorque}
    I\frac{d^2\theta_i}{dt^2}=\left[q\textbf{L}_{\hat{\mu}_i}\times\textbf{B}(x,y,z)+\sum_{j=1,j\neq i}^N{\frac{\mu_0}{4\pi}\frac{q^2}{|\textbf{r}_{\hat{\mu}_i,\hat{\mu}_j}|^3}(\textbf{L}_{\hat{\mu}_i}\times\textbf{r}_{\hat{\mu}_i,\hat{\mu}_j})}\right]\cdot(\hat{\mu}_i\times\hat{k})-\eta\frac{d\theta_i}{dt}
\end{equation}
\end{widetext}
where $I$ is the moment of inertia and $\theta_i$ is the angle of monopole $i$ relative to the $z$-axis. The first term in the right-hand side describes the torque due to a the magnetic flux density $\mathbf{B}(x,y,z)$ that, in general, is a function of space. We define $\textbf{L}_{\hat{\mu}_i}$ as the distance from the rotor to monopole $i$, which is free to rotate in the $\hat{\mu}_i$-$\hat{k}$ plane. The second term in the right-hand side describes the coupling between the monopoles $i$ and $j$ throughout the $N$ magnets in the array. Interactions between the same monopoles in the same magnet are not allowed. The distance between monopoles is $\textbf{r}_{\hat{\mu}_i,\hat{\mu}_j}=\textbf{L}_{\hat{\mu}_j}+\mathbf{d}_d-\textbf{L}_{\hat{\mu}_i}$, where $\mathbf{d}_l$ is the distance between the rotors of magnet $i$ and $j$. This implies that the distance varies in time as the magnets deviate from their equilibrium direction, providing a nonlinear coupling between the mechanical and magnetic forces. In general, the effective torque on the magnet is a three-dimensional vector. However, the magnets are constrained by the rotors so that only the $\hat{\mu}_i\times\hat{k}$ component of the torque leads to motion. Note that this cross product also ensures an appropriate definition of the torque sign in the Cartesian reference frame for magnets oriented along either the $x$ or $y$ axes. Finally, we also include energy dissipation through friction scaled by the parameter $\eta$.

The magnets utilized in the setup have diameter $D=1.59$~mm, length $2L=25.4$~mm, and mass $337\times10^{-6}$~kg. From these parameters, we calculate the magnet's moment of inertia for a cylinder $I=2.03\times 10^{-8}$~{kg m$^2$}. The saturation magnetisation was estimated to be $M_s=899$~kA/m, in good agreement with experimental determination with SQUID, see Supplementary Note~\ref{SI:SQUID}.

The solenoid used to drive the macro-ASI in experiments was modeled using the known expressions for the magnetic flux density due to a current loop with radial and normal-to-plane components
\begin{widetext}
\begin{subequations}
\label{eq:SIB}
\begin{eqnarray}
    \label{eq:SIBr}
    B_r&=&\frac{z}{r}\frac{B_0}{\pi\sqrt{(1+r/a)^2+(z/a)^2}}\frac{-K(m)+E(m)(1+(r/a)^2+(z/a)^2)}{(1+r/a)^2+(z/a)^2-4(r/a)}\\
\label{eq:SIBz}
    B_z&=&\frac{B_0}{\pi\sqrt{(1+r/a)^2+(z/a)^2}}\frac{K(m)+E(m)(1-(r/a)^2-(z/a)^2)}{(1+r/a)^2+(z/a)^2-4(r/a)}
\end{eqnarray}
\end{subequations}
\end{widetext}
where \(K(m)\) and \(E(m)\) are elliptical integrals of the first and second kind, with argument \(m=4r/a[(1+r/a)^2+(z/a)^2]\), and \(a\) is the radius of the solenoid. Because the magnets oscillate and their rotation angles change over time, the simulations measure the changes in the radial and vertical distances between the solenoid and each monopole at each time step. This allows for an accurate determination of the torque due to the external field at all times. We considered that the macro-ASI is located at a distance of $3.6$~cm over the solenoid and the measured radius of $a=10.5$~cm. We also consider that the solenoid is aligned with the geometrical centre of the macro-ASI. The flux density amplitude $B_0$ is a controllable parameter in the modeling. We calibrate the field by normalizing Eqs.~\eqref{eq:SIB} at the centre of the solenoid and offset $z$. This means that $r=z$. Numerical computation results in a normalization factor of $0.8895$.

Equation~\eqref{eq:SItorque} was solved using MATLAB's ode15s useful for stiff differential equations. We set outputs from the solver at increments of $0.01$ seconds and ran it for $50$~s to obtain a frequency resolution of $0.02$~Hz and a maximum frequency of $50$~Hz.

\section*{Acknowledgments}

This material is based upon work supported by the National Science Foundation under Grant No. 2205796 (L.S. and E.I.) and Grant No. DMR-2338060 (R.R.P. and D.A.B).

\section*{Author Contributions}

All authors contributed in the fabrication of the macro-ASI. R.R.P. and D.A.B. performed measurements and processed the data. L.S. and E.I. developed the numerical model. All authors analysed the data, discussed the results, and contributed to the manuscript.

\section*{Corresponding authors}
\noindent Correspondence to\\eiacocca@uccs.edu and \\ dbozhko@uccs.edu

\onecolumngrid
\clearpage

\textbf{\large{Supplementary material: Frequency comb in a macro-mechanical artificial spin ice}}
\\

Renju R. Peroor, Lawrence Scafuri, Dmytro A. Bozhko, and Ezio Iacocca

\setcounter{section}{0}
\renewcommand{\thesection}{SI \arabic{section}}

\setcounter{figure}{0}
\renewcommand{\thefigure}{SI \arabic{figure}}

\setcounter{table}{0}
\renewcommand{\thetable}{SI \Roman{table}}

\setcounter{equation}{0}
\renewcommand{\theequation}{SI \arabic{equation}}

\begin{figure}[h!]
\begin{center}
\includegraphics[width=5in]{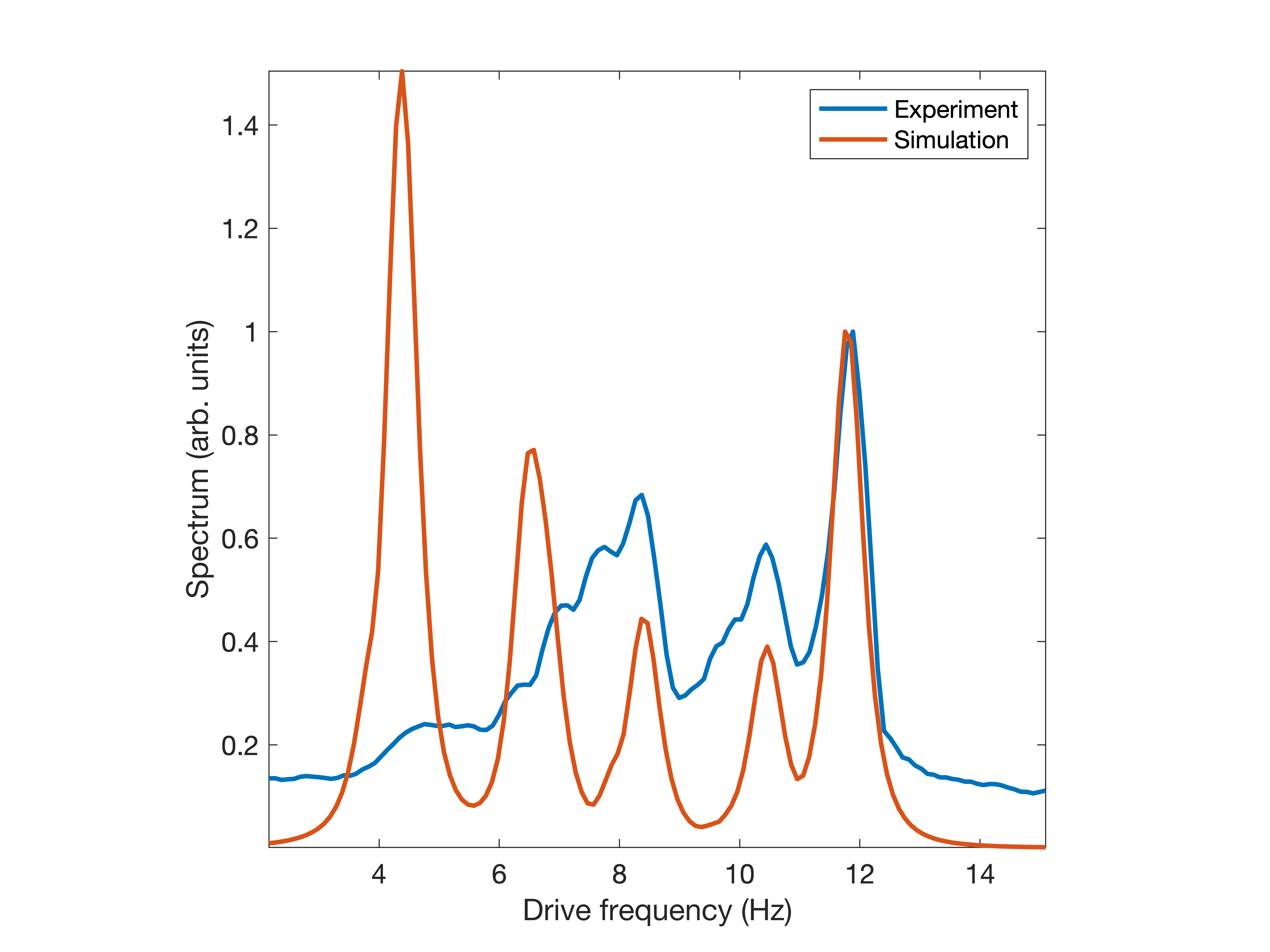}
\caption{\textbf{Experimental and numerical resonant modes.} The saturation magnetisation of the magnets was tuned in the model to match the highest-frequency mode. The magnitudes of the spectral peaks are normalized to coincide at the $12$~Hz mode. Excellent agreement is observed with modes at $\approx8.5$~Hz and $10.4$~Hz. The experiment shows a much richer spectrum in the range $6$~Hz to $10$~Hz. The simulations overestimate the low frequency mode, due to the ideal placement and conditions at the boundaries.}
\label{SI:fig1}
\end{center}
\end{figure}

\section{Squid}
\label{SI:SQUID}

To quantify the saturation magnetisation of the magnets we used a commercial Superconducting Quantum Interference Device (SQUID) magnetometer (Quantum Design MPMS2-XL). The hysteresis loops measurements were performed up to the maximal field of the setup of 5~T. A saturation magnetisation value of $863\pm15~$kA/m was obtained from the saturated magnetic moment $\mu$ using a simple relation $M_s=\mu\rho/M$, where $\rho$ is the density and $M$ is the mass. The density of the N42 magnetic material is reported by K\&J Magnetics \cite{KJMagnetics} to an average value of $7.45$~$\mathrm{g/cm^3}$. The mass of the measured sample was found to be $0.8$~mg using a precision balance.

\begin{figure}[h!]
\begin{center}
\includegraphics[width=6.5in]{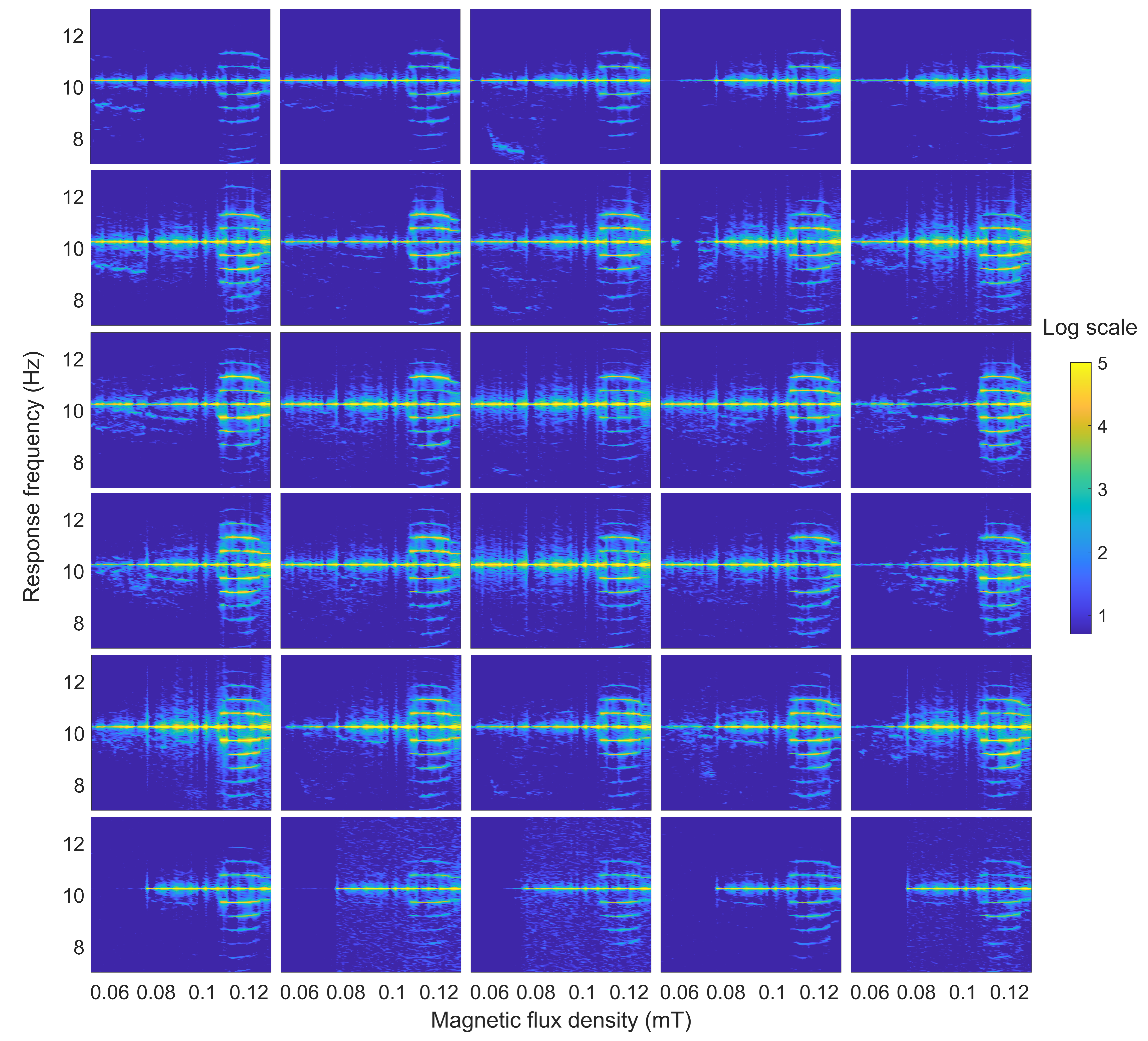}
\caption{\textbf{Frequency comb in array.} When the frequency comb is established, the dynamics are observed in all magnets. The above matrix show the magnetic flux density dependence for each magnet perpendicular to the camera. These are 30 magnets where the position of the spectra is correlated to their physical position. Clearly, the onset of the comb is observed in all magnets.}
\label{SI:fig2}
\end{center}
\end{figure}

\end{document}